\documentclass[english,prl, twocolumn]{revtex4}
\usepackage[T1]{fontenc}
\usepackage[latin9]{inputenc}
\usepackage{graphicx}
\usepackage{babel}
\usepackage{amssymb,amsmath, color}
\usepackage{subfigure}

\usepackage[usenames,dvipsnames]{xcolor} 

\begin{document}

\global\long\def\pgr{\mathcal{P}_{\text{gr}}}
\global\long\def\pdb{\mathcal{P}_{\text{db}}}
\global\long\def\pov{\mathcal{P}_{\text{ov}}}
\global\long\def\pn{\mathcal{P}_{0}}
\global\long\def\df{d_{\text{f}}}
\global\long\def\DCmax{G_1}
\global\long\def\Rv{\mathcal{R}_v}

\newcommand{\ie}{\emph{i.e.}}
\newcommand{\eg}{\emph{e.g.}}
\newcommand{\ER}{Erd\H{o}s-R\'{e}nyi}
\newcommand{\AP}{Achlioptas prcoesses}
\newcommand{\X}{\mathcal{O}}

\title{
{
Micro-transition cascades to percolation
}
} 

\author{Wei Chen$^{1}$, Malte Schr\"oder$^{2}$, Raissa M. D'Souza$^{3,4}$, Didier Sornette$^{5}$, Jan Nagler$^{5,2}$}
\affiliation{$^1$Institute of Computing Technology, Chinese Academy of Sciences, Beijing, China}
\affiliation{$^2$Max Planck Institute for Dynamics and Self-Organization (MPI DS) G\"ottingen, and
Institute for Nonlinear Dynamics, Faculty of Physics, University of G\"ottingen}
\affiliation{$^3$University of California, Davis, California 95616, USA}
\affiliation{$^4$Santa Fe Institute, 1399 Hyde Park Road, Santa Fe, New Mexico 87501, USA}
\affiliation{$^5$Department of Management, Technology and Economics (D-MTEC), ETH Zurich, Scheuchzerstrasse 7, CH-8092 Zurich, Switzerland}

\global\long\def\pcest{p_c^{\text{est}}}

\begin{abstract}

We report the discovery of a 
discrete hierarchy of micro-transitions occurring in models of continuous and discontinuous percolation.
The precursory micro-transitions
allow us to target almost deterministically the location of the transition point to global connectivity.
This extends to the class of intrinsically stochastic processes the possibility to use warning signals
 anticipating phase transitions in complex systems.

\end{abstract}

\maketitle

\paragraph{Introduction}

Percolation is a pervasive concept \cite{StaufferPercBook}, which has applications in a wide variety of natural, technological and social systems
\cite{Sornette1990, DrosselPRL1992,ParshaniandBulyrevandStanley2010,CallawayPRL2000,NewmanandWatts2002, saberi13},
ranging from conductivity of composite materials~\cite{Sahimi,AndradePRE2000} and polymerizations~\cite{ZiffPRL1982} to epidemic 
spreading~\cite{Anderson1991,CmoorePRE2000,PastorPRL2001} and 
information diffusion~\cite{StrangARS1998,Lazarsfeld1944}.
Across all percolation systems, once the density of links in the networked system exceeds a critical threshold the system undergoes a sudden usually unanticipated transition to global connectivity.

The prediction of tipping points and warning signals that precede a sudden transition have been a subject of high interest in many disciplines.
Generalized models, based on deterministic bifurcation dynamics, have been used to predict phase transitions triggered by small fluctuations \cite{Scheffler2009, Gore2012, BoettigerTheorecol2013, Gore2013, Boettinger2013}.
Here we report on a fundamental property of percolating systems which, in contrast, are dominated by (non-deterministic) large-scale disorder.

Discrete scale invariance (DSI) arises when the scale invariance of an observable $\mathcal{O}(x)\sim x^{\alpha}$ obeying $\frac{\mathcal{O}(\lambda x)}{\mathcal{O}(x)}=\lambda^{\alpha}$,
is broken such that the scaling relation does not hold for 
all  
$\lambda$ anymore but only for a countable set $\lambda_1,\lambda_2,...$
with a fixed $\lambda$ being the fundamental scaling ratio of the system and $\lambda_n=\lambda^n$ \cite{sornetteDSI, sornettebook}. 
Here, we unravel both genuine DSI and a generalized form of DSI in percolation, where in the latter the scaling ratio from the exponential is replaced by a scaling law. 
Analyzing individual events allows us to link these concepts.

Perhaps most importantly, we show that the emergence of global connectivity is announced by microscopic transitions of the largest component, 
the order parameter, well in advance of the phase transition.
{
We exemplify this for the generalized BFW model of genuinely
discontinuous percolation \cite{W.Chen and R. M. D'Souza, schrenk2012}, 
classic 
continuous percolation \cite{StaufferPercBook}, and globally competitive percolation \cite{nagler2011}. 
}
This suggests the universality of our findings.

\paragraph{Discontinuous percolation}

\begin{figure} 
\includegraphics[width=0.45\textwidth]{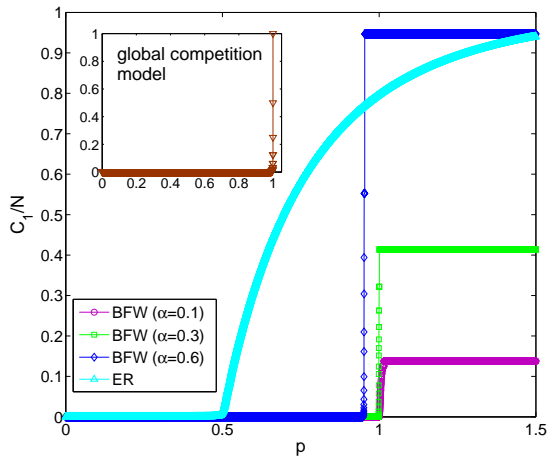}
\caption{
\label{BFW0} 
(Color online)
{\bf Discontinuous BFW percolation.} 
A typical realization of the relative size of the largest component $C_1/N$ as a function of link density $p$ for BFW with $\alpha=0.1,0.3,0.6$, 
and for the continuous ER model.
Inset: Discontinuous global competition model. 
System size $N=10^6$.
}
\end{figure}

The generalized Bohman-Frieze-Wormald model (BFW) is tailored to investigate discontinuous percolation transitions resulting from suppressing the growth of the largest component~\cite{W.Chen and R. M. D'Souza}, as characteristic of {\em explosive} percolation. 
The process is initialized with $N$ isolated nodes 
and a cap set to $k=2$ specifying the maximally allowed cluster size 
(a cluster is a set of linked nodes). 
Links are sampled one-at-a-time, uniformly at random from the complete network. 
If a link would lead to the formation of a component of size less than or equal to $k$ it is accepted. 
Otherwise, the link is rejected provided that the fraction of accepted links is greater than or equal to 
a function $g(k)=\alpha+(2k)^{-1/2}$, where $\alpha$ is a tunable parameter. 
Once rejecting a link would lead to the fraction of accepted edges dropping below $g(k)$, then 
$k\rightarrow k+1$ and the link is reexamined. This continues until either $k$ has increased sufficiently that the link can be accepted, or $g(k)$ becomes sufficiently small that
the link can be rejected. (See Supplementary Material \cite{Suppmat} for more details.)
Tuning the control parameter $\alpha$ allows for controlling the type and position of the phase transition,
as well as the number of giant components that abruptly emerge \cite{W.Chen and R. M. D'Souza,W.Chen and R. M. D'Souza arxiv}. 
Fig.~\ref{BFW0} shows the typical evolution of the relative size of the largest component $C_1/N$ 
as a function of the link density $p$ (\ie, number of links per node) for 
$\alpha=0.1,0.3,0.6$. 

The exact size of the largest component for a given link density may depend on the realization.
However, in traditional percolation at the thermodynamic limit
the order parameter, $C_1/N$, 
is believed to be globally continuous and thus not fluctuating---except at the phase transition points \cite{StaufferPercBook, aharony96, riordan, riordan2012}.
In contrast, we next demonstrate that the BFW model exhibits  peaks in the relative variance $\Rv$, well {\em before} the phase transition, which importantly do not disappear in the thermodynamic limit, and moreover, announce the phase transition. 
The relative variance of an order parameter $\X$, such as the total magnetization $\X=\mathcal{M}$, or the relative size of the largest component $\X=C_1/N$, 
is defined as
\begin{equation} \label{Rv}
 \Rv = \frac{\langle \X-\langle \X\rangle \rangle^2}{\langle \X\rangle^2} 
\end{equation}
where $\langle\rangle$ denotes ensemble averaging.

\paragraph{Micro-transition cascades to percolation}

\begin{figure} 
\includegraphics[width=0.45\textwidth]{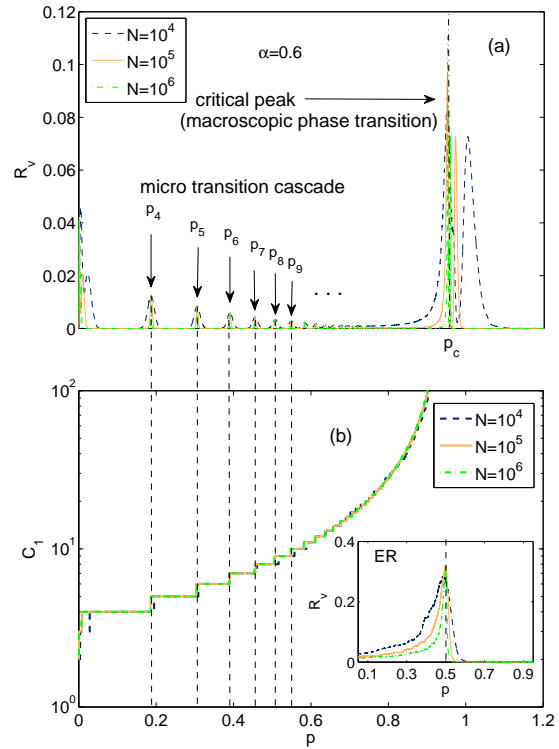}
\caption{
(Color online)
\label{BFW1}
{\bf Micro-transition cascade to percolation in the BFW model.}
(a) Relative variance $R_v$ versus link density $p$ showing 
sharp micro-transitions before $p_c$. 
Peaks after $p_c$ result from unstable giant components, discussed elsewhere \cite{chensouzanaglerpaper3}.
(b) The typical evolution (and collapse) of $C_1$
versus $p$, showing jumps 
when $C_1\rightarrow C_1+1$. 
Inset of (b): $\Rv$ versus $p$ 
for {continuous ER percolation,}
shown for three different system sizes. This reveals
 a {\em spectrum} of micro-resonances before $p_c=1/2$, that,
 in contrast to the BFW model, disappears as $N\rightarrow\infty$.
%
All data shown is the average over 1000 realizations.
}
\end{figure}

Fig.~\ref{BFW1}(a) shows sharp peaks in $\Rv$ well in advance of $p_c$
for the BFW model with $\alpha=0.6$ (figures for $\alpha=0.1$, $0.3$ are in the SI). 
This is unexpected as suggested from comparing 
Fig.~\ref{BFW1}(a), with the $\Rv$ plot for the 
 {
 \ER \ (ER) model  \cite{StaufferPercBook}
 }
 shown in the inset in Fig.~\ref{BFW1} (b).
In BFW we observe not only the standard transition to global connectivity, which is a 
{\em micro-macro-transition}, $C_1: o(N) \rightarrow O(N)$ at $p=p_c$,
but as well {\em micro-micro-transitions}, $C_1 \rightarrow C_1+1$  causing sharp jumps 
well before the emergence of global connectivity, Fig.~\ref{BFW1}(b).
{
Importantly, for increasing system size, the peaks become sharper, their positions converge to a well defined set,
and peak heights are independent of system size, see Fig.~\ref{BFW1}
and Supplementary Figs. S1-S5 \cite{Suppmat}.
}

We calculate the height of the $\Rv$ 
peaks, for jumps $C_1 \rightarrow C_1 + 1$, where the $i$-th jump corresponds to $C_1$ increasing from $i \rightarrow i+1$ at link density $p_i$. 
(The jump $1\rightarrow 2$ occurs always when the first link is added, thus no peak of $R_v$ is observed then.)
We estimate the maximum of $R_v$ for the $i$-th jump by assuming that for a fraction $q_i$ of the realizations $C_1\rightarrow C_1+1$, while $C_1$ for a fraction $1-q_i$ of the realizations has not increased. Hence, from 
Eq.~(\ref{Rv}) 
we obtain 
\begin{equation} \label{Rvheight}
 \Rv(p_i) = \frac{q_i(1-q_i)}{(i+q_i)^2} \ \text{with} \ q_i = \frac{i}{2i+1},
\end{equation}
where the $q_i$'s satisfy $\frac{\partial{\Rv}}{\partial{q_i}}=0$.
From Eq.~\ref{Rvheight}, we find $q_4=4/9$ 
$q_5=5/11$, and
$ q_6=6/13$, and that
$\Rv(p_4)\approx0.0125, \Rv(p_5)\approx 0.0083, \Rv(p_6)\approx0.0060$ for the $o(N)$-transitions $4\rightarrow 5, 5\rightarrow 6$, and $6\rightarrow 7$, respectively.
{
These predictions are well supported by numerics, 
see \cite{Suppmat}. 
}

\begin{figure} 
\includegraphics[width=0.45\textwidth]{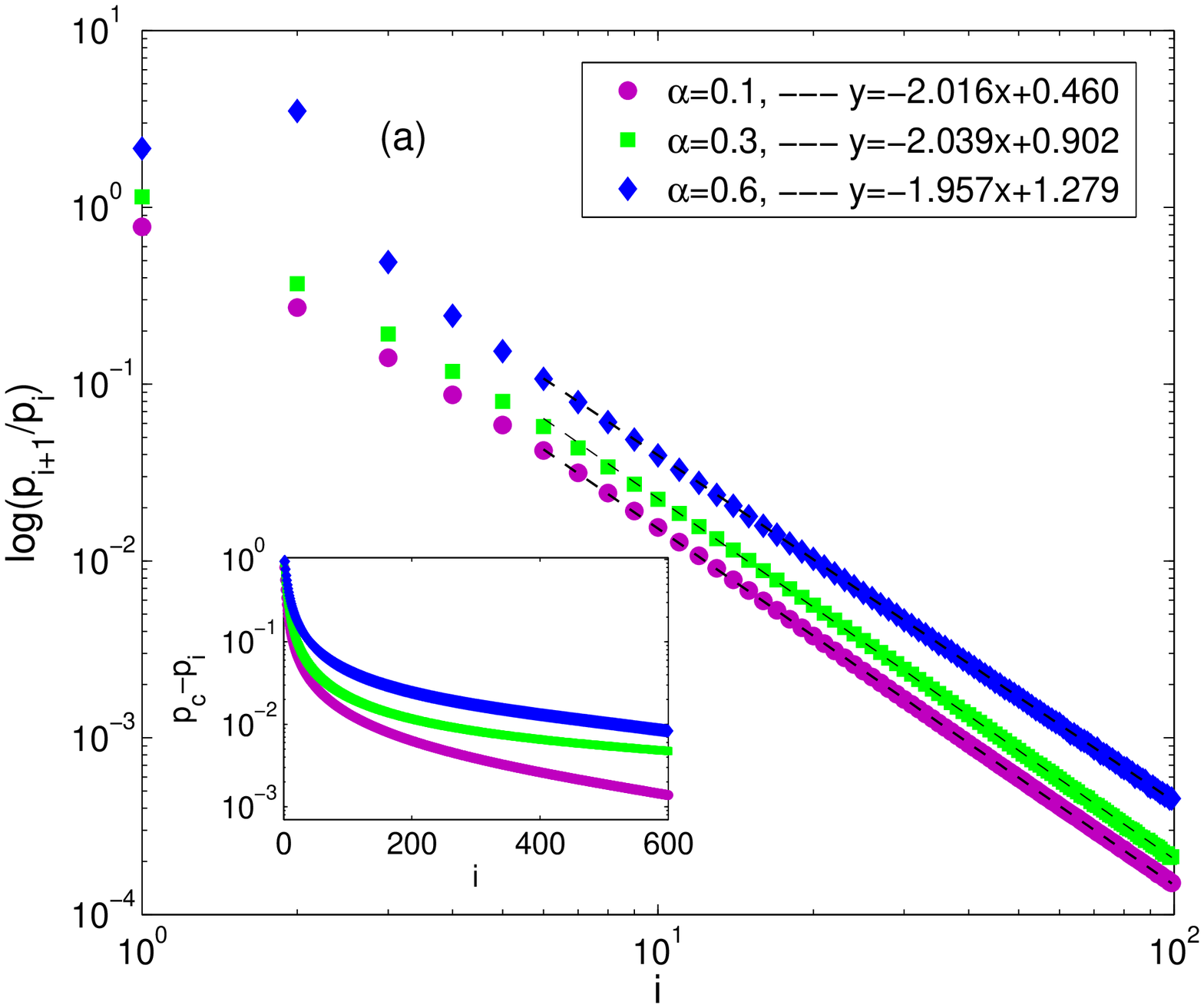}
\includegraphics[width=0.45\textwidth]{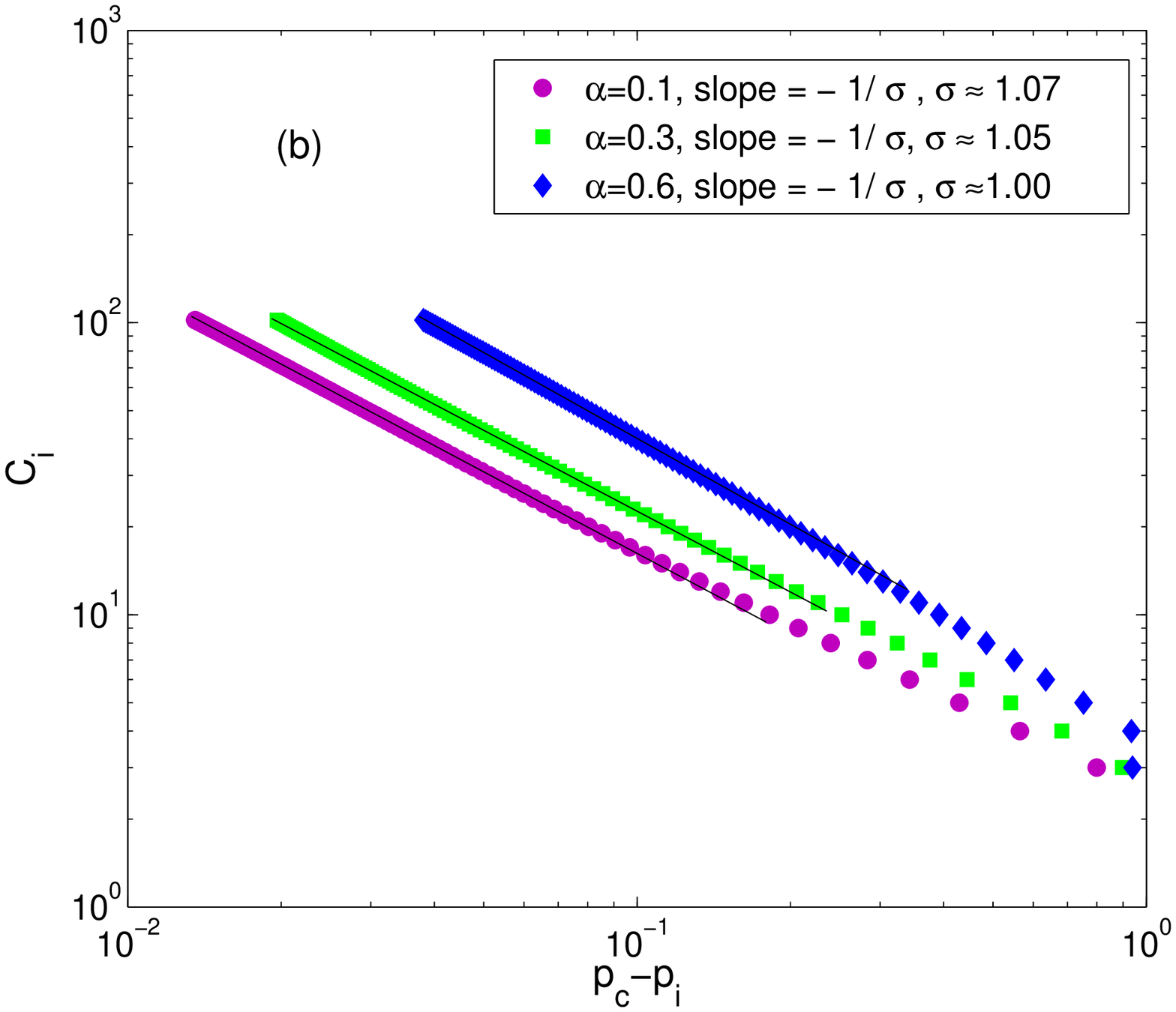}
\caption{
\label{BFW3}
(Color online)
{\bf Scaling Laws and Convergence to $p_c$ for the BFW model.}
 (a) The position $p_i$ of the micro-transitions are well fitted by Eq.~\ref{pis} 
 for $\alpha=0.1,0.3,0.6$.
  Note that we display a log-log plot suggesting $\log{\frac{p_{i+1}}{p_i}}=A i^{-b}$.
 Inset of (a): Evidence for $p_i \rightarrow p_c$ for $\alpha=0.1,0.3,0.6$ and $N=10^7$.
(b) Exponent $\sigma$ defined in Eq.\ (\ref{PT2}) for $\alpha=0.1, 0.3, 0.6$ and $N=10^7$. 
}
\end{figure}

Analyzing additional peaks as shown in Fig.\ \ref{BFW3}(a) suggests a scaling law of the relative peak positions
\begin{equation} \label{pis}
 \frac{p_{i+1}-p_i}{p_i} \approx \log\left(\frac{p_{i+1}}{p_i}\right) = A i^{-b}, \;i\gg 1
\end{equation}
with $b$ close to 2, slightly depending on $\alpha$, for some $A>0$. 

We infer $p_\infty$ from Eq.~\ref{pis} 
(See \cite{Suppmat} for details) and find that $p_\infty=p_c = 0.940, 0.998, 0.999$ for $\alpha = 0.6, 0.3, 0.1$ respectively, which agree exactly with the values of $p_c$ obtained from direct simulation of the BFW model. 
(See \cite{Suppmat} for $p_\infty$ values obtained for additional $\alpha$ values.)
In fact, the inset of Fig.~\ref{BFW3}(a)
shows that $p_c-p_i<0.01$ when $i>600$ for $\alpha=0.1,0.3,0.6$.
Thus we find here that the positions of the micro-transitions announce the phase transition.

\paragraph{Discrete scale invariance in percolation}

Next 
we show that a percolation model with global competition for link-addition
exhibits 
a discrete scale invariance that underlies the observed cascade to percolation. 

Start with $N$ isolated nodes.
At each step connect the two smallest clusters in the system (if there are multiple choices, throw a fair dice to choose among the equivalent cluster pairs)  \cite{friedman2009, nagler2011}. 
In this model all possible links compete for addition.
Thus 
it is the limiting case $m\rightarrow\infty$ of the original explosive percolation models from Ref. \cite{achlioptas}, where at each step a fixed number of $m$ links compete for addition \cite{friedman2009, nagler2011}.
The global competition suppresses transitions different from {\em doubling} transitions 
$C_1\rightarrow 2 C_1$ resulting in $p_c=1$.
For $N\gg 1$ fixed, these occur at $p_n=\frac{2^n-1}{2^n}$, $n$ integer \cite{nagler2011},
and hence
\begin{equation}\label{GCpi}
 p_{n} = p_c -2^{-n}, \; n\ge 0
\end{equation}
%
As a result, the doubling transitions announce the percolation transition as $p_n \rightarrow p_c$ for $n\rightarrow \infty$. 
This is 
a signature of discrete scale invariance (DSI) \cite{sornetteDSI,sornettexx} 
as we can rewrite Eq.\ (\ref{GCpi}) to
{
\begin{equation}\label{GCDSI}
 \frac{p_c-p_{n+1}}{p_c-p_n} = 1/\lambda, \; C_1(p_{n+1})=\lambda C_1(p_n)
\end{equation}
with the discrete scaling factor $\lambda=2$.
}

The DSI can be broken when the system stochastically deviates from the strict size doubling rule, as generically given in percolation and other disordered systems \cite{sornettebook}.
{
We thus consider jumps from any size $C_1\le i$ to precisely $C_1=i+1$. 
The index transformation $i=2^{n+1}-1$ formally breaks the genuine DSI and
suggests, using Eq.(\ref{GCpi}), the transition positions
%
 $p_i = 
 1-2/(i+1) 
 $.
%
}

The relative positions of the transitions then read
{
\begin{equation}\label{eq:BDSI}
\frac{p_{i+1}-p_i}{p_i}  
\sim  i^{-b}, \text{for} \; i\gg 1,
\end{equation}
}
with $b=2$,
which agrees well with the scaling law Eq.\ (\ref{pis}). 
It is easy to see that any transformation of type $n \rightarrow \alpha \log(\beta i+\gamma)$, with constants $\alpha,\beta>0$ and any $\gamma$ gives the same qualitative result.

\paragraph{Relation to cut-off critical exponent}

Next we demonstrate that micro-transitions also announce the phase transition well in advance for continuous percolation.
In continuous percolation as $p\rightarrow p_c$, from below ($p<p_c$), the emergence of the giant cluster
 is characterized by
\begin{equation}\label{PT1}
 C_1 \sim  (p_c-p)^{-\frac{1}{\sigma}}
\end{equation}
where $\sigma$ is the cut-off critical exponent that, given strong disorder, 
is related to the correlation exponent $\nu$ and the fractal dimension $d_f$ via $\sigma=\frac{1}{\nu d_f}$ \cite{stanley, StaufferPercBook}.

\begin{figure}
\includegraphics[width=0.45\textwidth]{
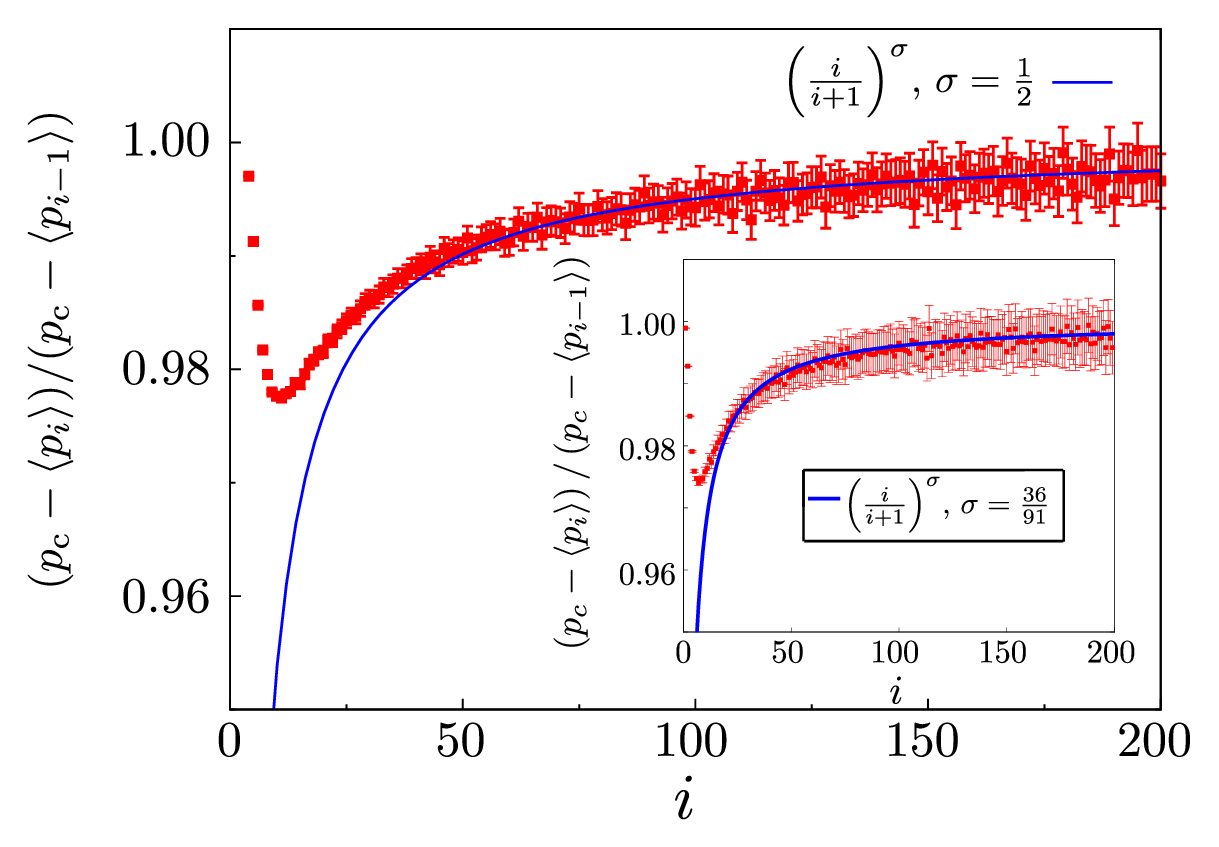}
\caption{ 
\label{fig:ER}
(Color online)
{\bf Scaling relations for continuous percolation.}
Numerical evidence for the prediction
$\frac{p_c-p_{i}}{p_c-p_{i-1}} \sim (\frac{i}{i+1})^{\sigma}$ from Eq.\ (\ref{PT2}),
for ER ($\sigma=1/2$, $N=2^{25}$, 30000 realizations) 
and 2D lattice ($\sigma=36/91$, $N=1024\times 1024$, 30000 realizations).
%
}
\end{figure}

We estimate the positions of the micro-transitions at $p_i$ from (\ref{PT1}) for $C_1=i+1$ and $p=p_i$.
Solving for $p_i$ gives {
\begin{equation}\label{PT2}
 p_i  = p_c - A (i+1)^{-\sigma}
\end{equation}
}
with some prefactor $A>0$.
From Eq.\ (\ref{PT2}) we find
\begin{equation} \label{PT3}
 \frac{p_{i+1}-p_i}{p_i}
 {
  \approx \frac{A[(i+1)^{-\sigma} - (i+2)^{-\sigma}]}{p_c}
  }
   \sim i^{-(1+\sigma)} \; \text{for} \;i\gg 1.
\end{equation}
{
This equation predicts for {\em any} phase transition characterized by the exponent $\sigma$ 
a cascade defined by Eqs. (\ref{eq:BDSI}) and (\ref{PT3}) 
with exponent $b=1+\sigma$. 
}

Above the percolation upper critical dimension, and thus 
for ER percolation, 
the set of critical percolation exponents are known, $\sigma=\nu=1/2$, $d_f=4$ \cite{StaufferPercBook}.
%
For ER, Eq.\ (\ref{PT2}) is well supported by numerics, see Fig.\ \ref{fig:ER}.
Further,
 numerics for 2D site-percolation, where 
 $\sigma=\frac{1}{\nu d_f}=\frac{1}{4/3 \ 91/48}\approx0.396$
 is known from theory \cite{StaufferPercBook},
 well supports our prediction (see inset in Fig.\ \ref{fig:ER}).

Specifically,
we define $p_i$ as the position of the micro-transition of the type $C_1: x\rightarrow i+1\; (x\leq i)$.
Since for a given realization a jump of this type and thus $p_i$ may not exist,
 to obtain $\langle p_i\rangle$ in Fig.\ \ref{fig:ER},  we average for each $i$
 over all realizations where $C_1: x\rightarrow i+1\; (x\leq i)$  do occur and $p_i$ is well defined,
 \begin{equation}\label{eq:avpi}
 \langle p_i \rangle := \langle  \arg_{\exists (i,x\le i)}\{ C_1(p):x\rightarrow i+1\}  \rangle
 \end{equation}
In contrast, for fixed $N$, most pronounced close to the origin at $p=0$, micro-transitions localize
but ensemble averaging 'blurs out' peaks in $\Rv$ for larger values of $p$ (see 
Figs. S7 and S8 \cite{Suppmat}). 

%
For the BFW model we find the exponent $\sigma$, slightly depending on $\alpha$, close to unity, see Fig.\ \ref{BFW3}(b).
This result is 
in agreement with Eq.~(\ref{eq:BDSI}), predicting $b\approx 2$, and with the numerics shown in Fig.\ \ref{BFW3}(a).

For the globally competitive percolation model we calculate for $p< p_c$ \cite{nagler2011,nagler2013}
\begin{equation}\label{GSIGMA}
 C_1 = \frac{N}{N-L}=\frac{1}{1-p}=(p_c-p)^{-\frac{1}{\sigma}}, \ p=L/N, 
\end{equation}
with $\sigma=1$, which is an exact result. 

Further, from Eq.\ (\ref{PT1}) we calculate the relative positions for transitions of type $C_1\rightarrow n C_1$, for $n>1$ fixed,
%
%
\begin{equation} \label{PT4}
 \frac{p_{c}-p_{ni}}{p_c-p_i} \rightarrow n^{-\sigma}=n^{1-b}=:1/\lambda^{(n)}, \ \text{for}\ i\rightarrow \infty.
\end{equation}
%
%
Eq.\ (\ref{PT4}) describes a family of micro-transition scaling relations parametrized by $n$.

We can also turn Eq.\ (\ref{PT4}) around for predicting $p_c$.
For the ER model we find for $n=2$, $\lambda^{(2)}=2^{\sigma}=\sqrt{2}$
and 
\begin{equation}\label{eq:pc}
 p_c = \lim_{i\rightarrow\infty} \frac{\lambda^{(2)} p_{2i}-p_i}{\lambda^{(2)}-1}.
\end{equation}
%
Numerical evaluation of (\ref{eq:pc}) suggest $p_c=0.499585$ for $i=128$, which is close to the exact value $p_c=0.5$~\cite{StaufferPercBook}.

\paragraph{Conclusion}

We have established the appearance of well defined peaks in the subcritical regime for standard processes of continuous and discontinuous percolation.
The cause of those resonances in the relative fluctuation function are micro-transitions of type $o(N)\rightarrow o(N)$ that generically announce the percolation phase transition well in advance of $p_c$.
Therefore, genuine peaks in the relative variance do not necessarily indicate a phase transition point, as it is commonly exploited for characterization of the phase transition point
in classical and quantum critical systems \cite{stanley}.
We have discovered an overlooked phenomenon, micro-transition cascades in percolation, 
which as shown here can
result from a (generalized) discrete scale invariance of the order parameter at and before criticality.

Globally competitive percolation displays genuine discrete scale invariance where 
the positions of the micro-transitions are characterized by powers of the single fundamental scaling factor $\lambda=2$.
This results from a single route of doubling transitions of the order parameter, for large finite systems.

In contrast,
we have demonstrated that
systems with strong disorder display multiple micro-transition cascades to percolation that are not characterized by a single scaling factor
but by a set of scaling relations, exemplified for percolation. The simplest subset of these scaling relations describe transitions $C_1\rightarrow n C_1$, $n\ge2$ integer, which occur
at localized positions, for large finite systems.  We call this phenomenon generalized discrete scale invariance in percolation.

We have established a novel type of finite size scaling laws which crucially characterize percolation. 
As our arguments are independent of the percolation process and the system size, for any $N<\infty$
there necessarily exist cascades to percolation imprinted  both in the order parameter and its relative variance.
Exemplified for a well-studied discontinuous percolation process,  we have shown that these cascades can even survive the thermodynamic limit.

Continuous percolation exhibits a continuous power law divergence at $p_c$ that does not show any localized 
peaks in the relative variance in the thermodynamic limit.
In contrast, for fixed $N$, 
micro-transitions do localize albeit ensemble averaging blurs out peaks in $\Rv$. 
%
%
Ensemble averaging  in accords with Eq.\ \ref{eq:avpi}, however,
 robustly unravels the discrete hierarchy 
and thus overcomes the effect of blurring.

We find DSI and its (exponential or power law) scaling laws from a non-trivial exponentiation
 ($C_1\rightarrow \lambda C_1$ at $p_i\rightarrow p_{i+1}$) of a
discrete translational invariance resulting from the discreteness of the network, or lattice \cite{sornetteDSI}. 

Hence, a
percolation phase transition can be anticipated by inferring information
from ensemble averaged microscopic state changes of the order parameter well in advance of the transition point.
Thus we are able to extend the possibility of early warning signals to classes of stochastic dynamics.
Future work must establish if these findings will open new avenues for the prediction of phase transitions unrelated to percolation.

\begin{acknowledgements}

This work was funded by the 973 National Basic Research Program of China under grant numbers 2013CB329602 and 2012CB316303, the National Natural Science Foundation of China under grant numbers 61232010 and 11305219, National High Technology Research and Development Program of China under grant number 2014AA015103,
the Defense Threat Reduction Agency HDTRA1-10-1-0088, and the Army Research Office awards W911NF-09-2-0053 and W911NF-13-1-0340. 

\end{acknowledgements}

\end{document}


\title{{\Large {\it Supplementary Information:}\\ Micro-transition cascades to percolation}}

\author{Wei Chen, Malte Schr\"oder, Raissa M. D'Souza, Didier Sornette, Jan Nagler}

\date{\empty}
\maketitle

\vspace{-0.1in}

Here we provide supplementary information to the main manuscript.

\section{Additional notes to the notion of discrete scale invariance (DSI)}

DSI is the partial breaking of continuous scale invariance \cite{stanley, sornettebook}
in the sense that the scale invariance only holds for magnification factors (or scaling factor)
that are integer powers of a fundamental scaling ratio. For instance, the standard Cantor set defined as a segment divided in 3 parts
in which the central segment is deleted, and the operation is repeated ad infinitum, exhibits the property of DSI
 with a fundamental scaling ratio of 3 \cite{sornetteDSI}.\\
 
The key idea here is that it is a kind of exponentiation of the discrete increase $C_1\rightarrow C_1+1$ (or $C_1\rightarrow C_1+n$) that leads to DSI and its generalization ($e^{C_1}\rightarrow e^{C1} * e^{a*1}$, where $a$ depends on the physical or geometrical problem). Hence the scaling ratio $\lambda$ comes from the transformation by a kind of exponentiation of the discrete 
translational invariance of a discrete growing process.
In particular, our micro-transitions can be seen as macro-evidence of the existence of an underlying discrete lattice or process combined with the cooperation / connectivity associated with the percolation rules. \\

The simplest case of the exponentiation of DSI occurs in the global competition percolation model  
and is described by $C_1 \rightarrow \lambda C_1$ such that
$C_1^{(n)}=\exp(\log(\lambda)n)C_1(n_0)$, $\lambda=\text{const.}=2$, $n=1,2,...$ (see Eq. (5) in the main text).
Thus we find discrete jumps of an extremal stochastic variable (here the order parameter $C_1$)
whose micro-transition positions $p_n$ (or $p_i$) are described by scaling laws characterizing the translational invariance of the system,
or analogously, the waiting times between extremal growth events (in $C_1$).
In general, this genuine DSI is broken such that the exponential of $\log(\lambda)n$ is replaced by a more complex expression
such as power law scaling relations (Eqs. (5) and (11) in the main text). \\

The order parameter of continuous percolation becomes perfectly smooth in the limit $N\rightarrow\infty$, exhibiting a typical power law divergence
of the order parameter at $p_c$. DSI relates to finite size ($N<\infty$) scaling laws in $p_i$ which robustly announce $p_c$.
As being evident from our study the announcement begins well in advance of $p_c$ for both the continuous and discontinuous models which is thus
unrelated to the standard (dis)continuous divergence close to $p_c$.
In order to be able to make a general argument we focus on the characterizations of the $p_i$ close to $p_c$ where we find clean
scaling laws which then connect to the critical cut-off exponent $\sigma$.\\

To summarize, we find DSI and its (exponential or power law) scaling laws from a non-trivial exponentiation of a
discrete translational invariance of the positions of micro-transitions of an extremal stochastic variable (the order parameter $C_1$).

\section{Connection to Lattice Animals}

We have demonstrated that micro-transitions occur in four different percolation models, 
which  shows that they are not an artifact of a specific construction method. Instead, they actually result
from the discrete nature of the lattice or network, in a kind of
exponentiation of the translational invariance of integer increase to
(discrete) scale invariance across scales.
This is in agreement with the finding that lattice animals in percolation
do show discrete scale invariance \cite{saleur}. The novelty here is that the information
on $p_c$ is already contained in the discrete scaling of the small clusters.

\section{The Bohman-Frieze-Wormald algorithm}
Stating the BFW model explicitly, it begins with a collection of $N$ isolated nodes and proceeds in phases starting with $k=2$, which is the maximally allowed cluster size. Edges are sampled one at a time, uniformly at random from the complete graph induced on the $N$ nodes. 
Using the notation in~\cite{T. Bohman and A. Frieze}, let $u$ denote the total number of edges sampled, $A$ the set of accepted edges (initially $A =\emptyset$), and $t=|A|$ the number of accepted edges. At each step $u$, the selected edge $e_u$ is examined via the following algorithm (the BFW algorithm):
 
\vspace{2mm} 
{\tt Set $l=$ maximum size component in $A \cup \{e_{u}\}$

\ \ \ if $\left(l\leq k\right) \{$
 
\ \ \ \ \ \ $A \leftarrow A \cup \{e_{u}\}$
 
 \ \ \ \ \ \ $u\leftarrow u+1 \ \}$
  
\ \ \ else if $\left(t  / u < g(k)\right) \{$
 $k\leftarrow k+1\ \}$
 
\ \ \  else $\{\ u \leftarrow u+1 \ \}$

}

\vspace{2mm}
\noindent
Walking through the loop, if adding $e_u$ leads to a component of size $l \le k$, the edge is added and step $u$ completes and we move onto step $u+1$.  Otherwise, check if $t/u < g(k)$.  If not, then $t/u \ge g(k)$ and $e_u$ is ignored ({\it i.e.} rejected) and we move onto step $u+1$.  If instead it is the case that $t/u < g(k)$ then we augment $k$ and walk through from the ``if" loop from the start again. 
In other words, while $t/u < g(k)$, $k$ is augmented by one repeatedly until either  $k$ becomes large enough that edge $e_u$ is accepted or $g(k)$ decreases sufficiently that edge $e_u$ can be rejected at which point step $u$ finally ends.  Note $g(k)=1$ requires that all edges be accepted, equivalent to \ER~\cite{ER}.  Also we use the ``$\min$" function, $g(k)=\min\left(1,1/2+({2k})^{-\beta}\right)$, to ensure $g(k) \le 1$ for all choices of $\beta$, since the fraction of nodes accepted cannot exceed unity. 

T. Bohman, A. Frieze, and N. C. Wormald established rigorous results whereby setting $g(200)=1/2$, all components are no larger than $k=200$ nodes (\ie, no giant component exists) when $m=0.96689N$ edges out of $2m$ sequentially sampled random edges have been added to graph \cite{T. Bohman and A. Frieze}. They further establish that a giant component must exist by the time $m=c^*N$ out of $2m$  sampled edges have been added,  with $c^{*} \in [0.9792,0.9793]$.  Yet, they did not analyze the nature of the percolation transition.
  
In~\cite{W.Chen and R. M. D'Souza} it was shown that the original BFW model leads to the simultaneous emergence of two giant components (each with fractional size $C > 0.4$), and the stability of the multiple giants was analyzed.  (Essentially, once in the supercritical regime, there are always sufficient edges internal to components sampled that whenever an edge connecting two giant components is sampled it can be rejected.) It was also shown in~\cite{W.Chen and R. M. D'Souza} that by using the function $g(k) = \alpha + (2k)^{-1/2}$ , then the parameter $\alpha$ tunes the number of stable giant components that emerge simultaneously. For suitable choice of $\alpha$, a second continuous transition appears in the supercritical regime, leading to the emergence of a second giant component after the percolation threshold~\cite{ChenNaglerDSouza2013}. The underlying mechanism for the discontinuous transition occurring in the BFW model is the domination of the overtaking mechanism in the growth of the largest component~\cite{ChenZhengDSouza}.

 \section{Micro-transition cascade to percolation}
\begin{figure}

 \centering
 \includegraphics[width=0.8\textwidth]{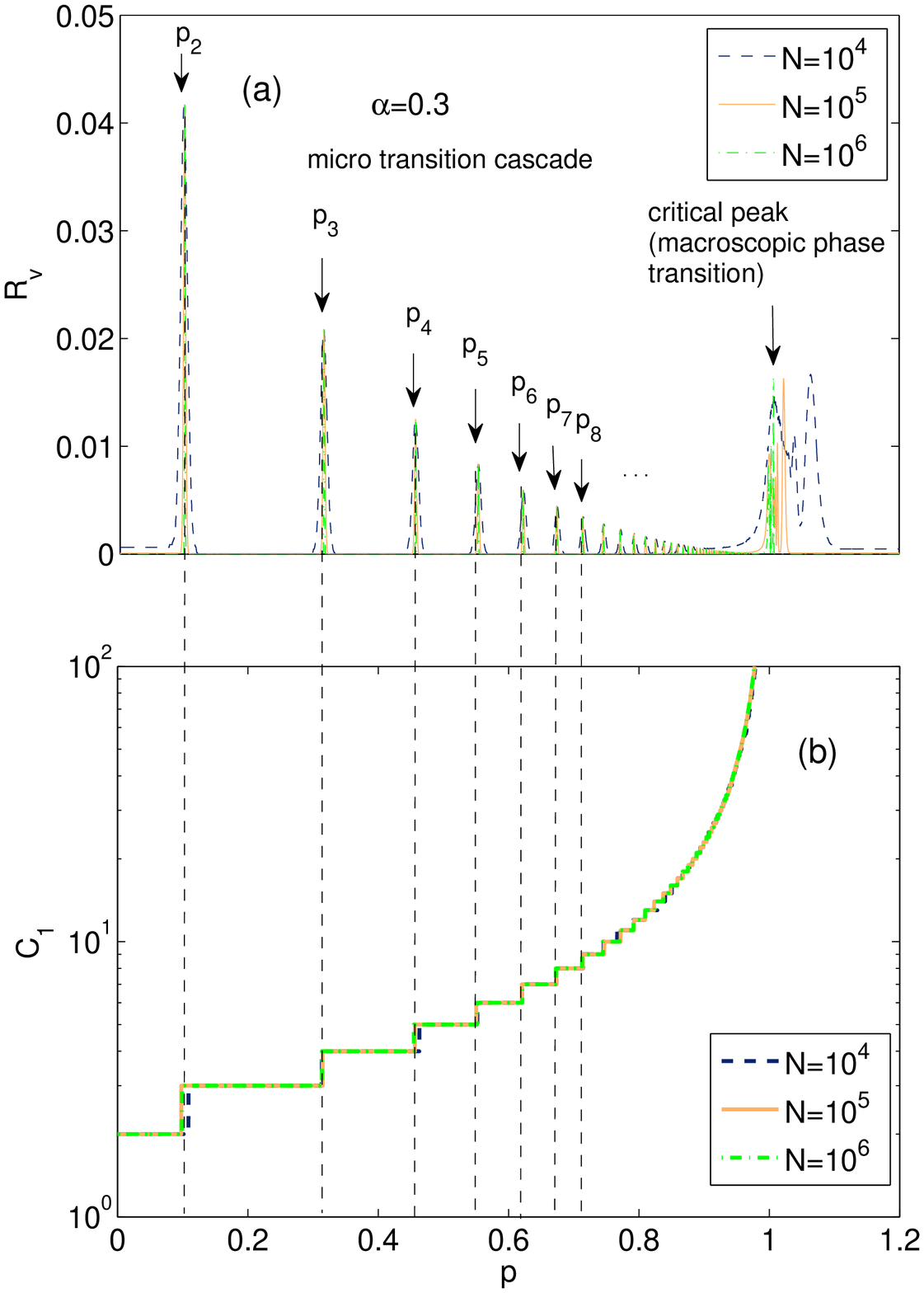}
 \vspace{-0.2in}
\caption{BFW model with $\alpha=0.3$: (a) relative variance $R_v$ versus edge density $p$. Infinite sharp micro-transitions before $p_c$ and multiple peaks at $p_c$. (b) The typical evolution of $C_1$ versus $p$, showing the microscopic jumps $C_1\rightarrow C_1+1$ occurs at the peaks of $R_v$ before $p_c$.}
 \label{fig:as}
 \end{figure}

\begin{figure}

 \centering
 \includegraphics[width=0.8\textwidth]{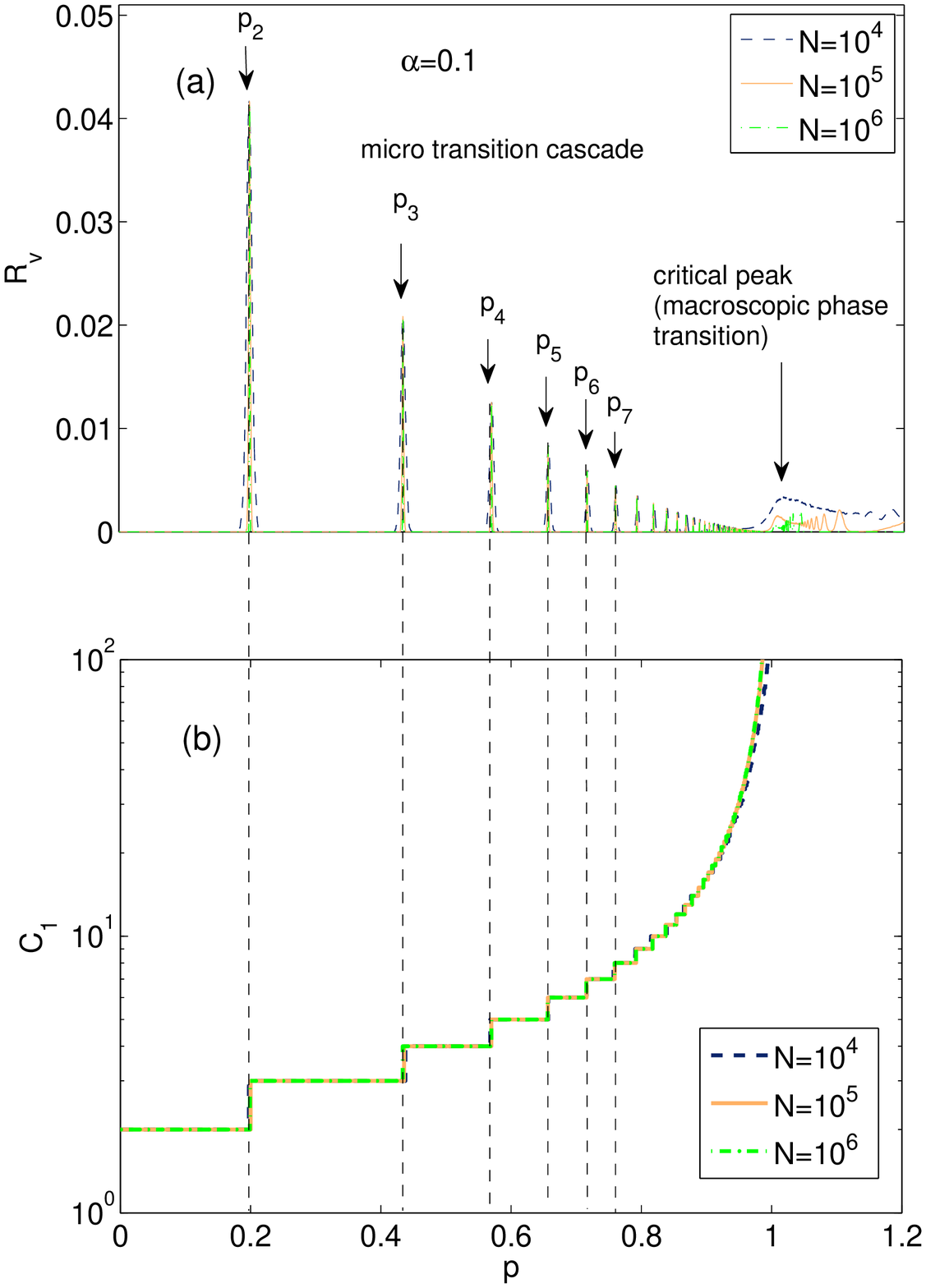}
 \vspace{-0.2in}
\caption{BFW model with $\alpha=0.1$: (a) relative variance $R_v$ versus edge density $p$. Infinite sharp micro-transitions before $p_c$ and multiple peaks at $p_c$. (b) The typical evolution of $C_1$ versus $p$, showing the microscopic jumps $C_1\rightarrow C_1+1$ occurs at the peaks of $R_v$ before $p_c$.}
 \label{fig:bs}
 \end{figure}

\begin{figure} 

\centering
\includegraphics[width=0.8\textwidth]{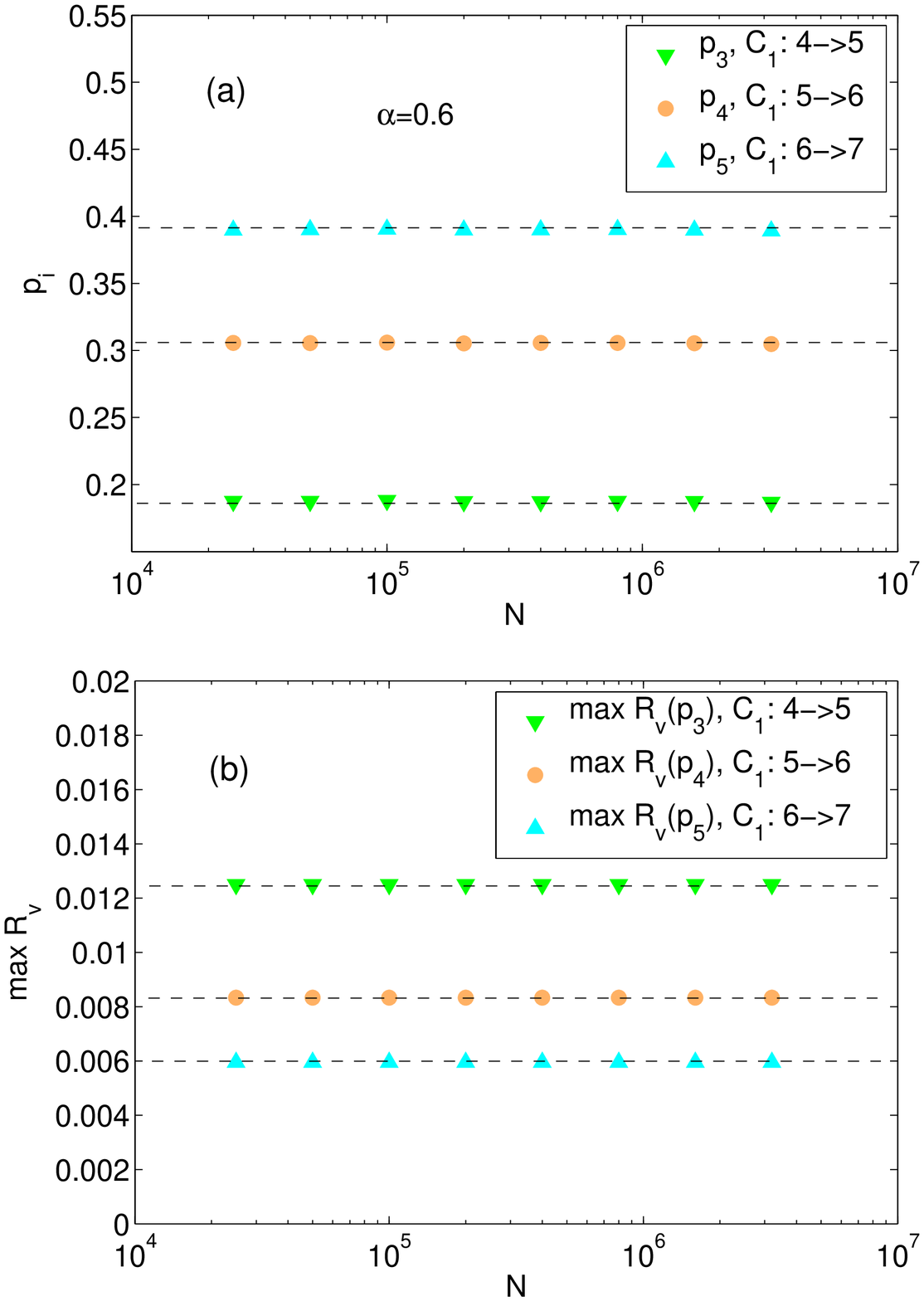}
\caption{
Localization of the peaks for the BFW model with $\alpha=0.6$:
(a) The location of the $i$-th peaks of $\Rv$, $p_i, i=3,4,5$ is independent of the system size $N$ for BFW model with $\alpha=0.6$. 
(b) The heights of the $i$-th peaks of $\Rv$, $\Rv(p_4), \Rv(p_5), \Rv(p_6)$ are independent of the system size $N$.  
$\Rv(p_i)$ with $i>6$ are also independent of the system size $N$, which are not shown for space constraints.
\label{fig:BFW2}
}
\end{figure}

\begin{figure} 

 \centering
 \includegraphics[width=0.8\textwidth]{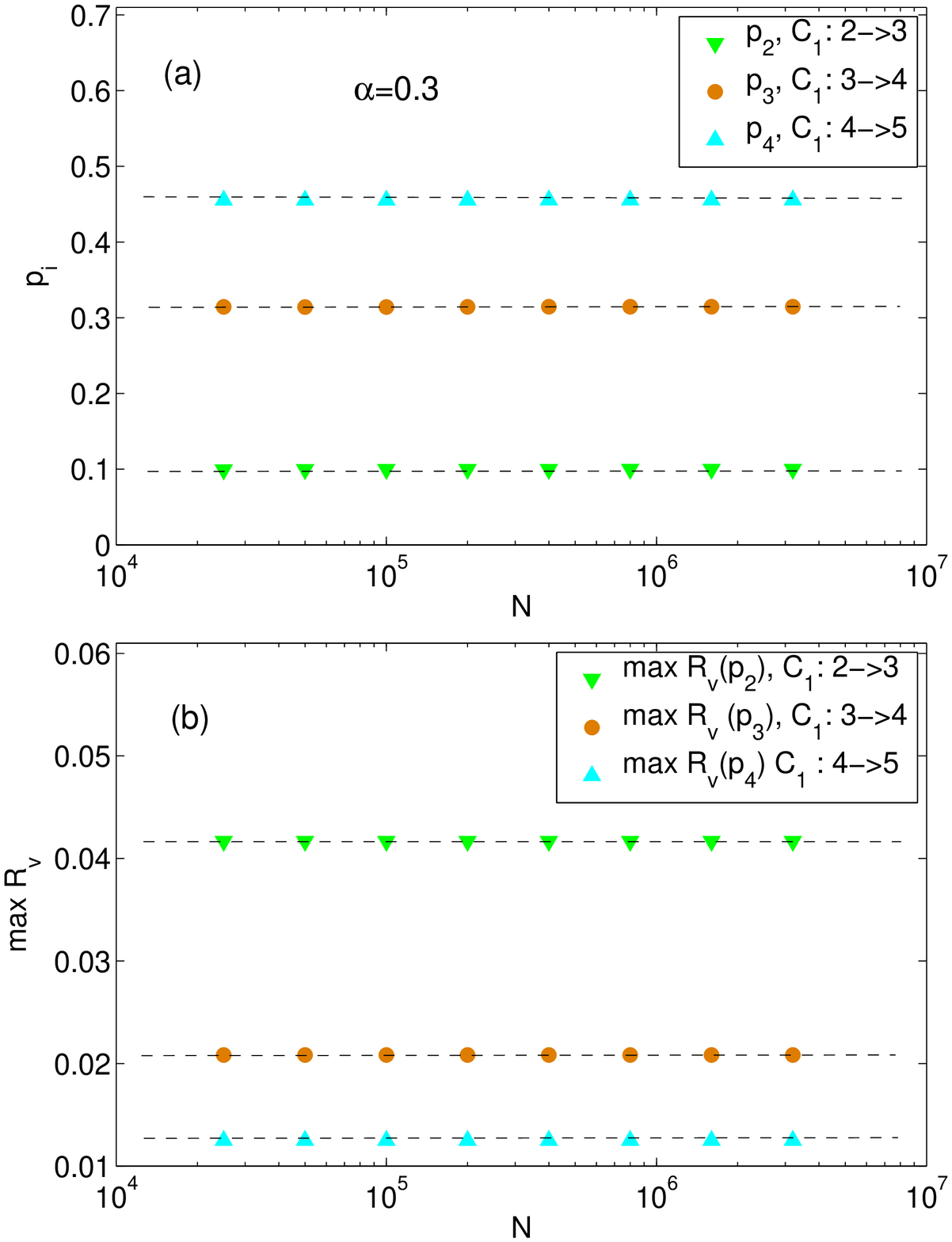}
 \vspace{-0.2in}
\caption{BFW model with $\alpha=0.3$: (a) the location of the $i-th$ peaks of $R_v$, $p_i, i = 2, 3, 4$ is independent of the system size $N$ for BFW model with $\alpha=0.3$. (b) the heights of the $i-th$ peaks of $R_v$, $R_v(p_2), R_v(p_3), R_v(p_4)$ are independent of system size $N$. $R_v(p_i)$ with $i>4$ are also independent of the system size $N$, which are not shown for space constraints.
}
 \label{fig:cs}
 \end{figure}

\begin{figure}

 \centering
 \includegraphics[width=0.8\textwidth]{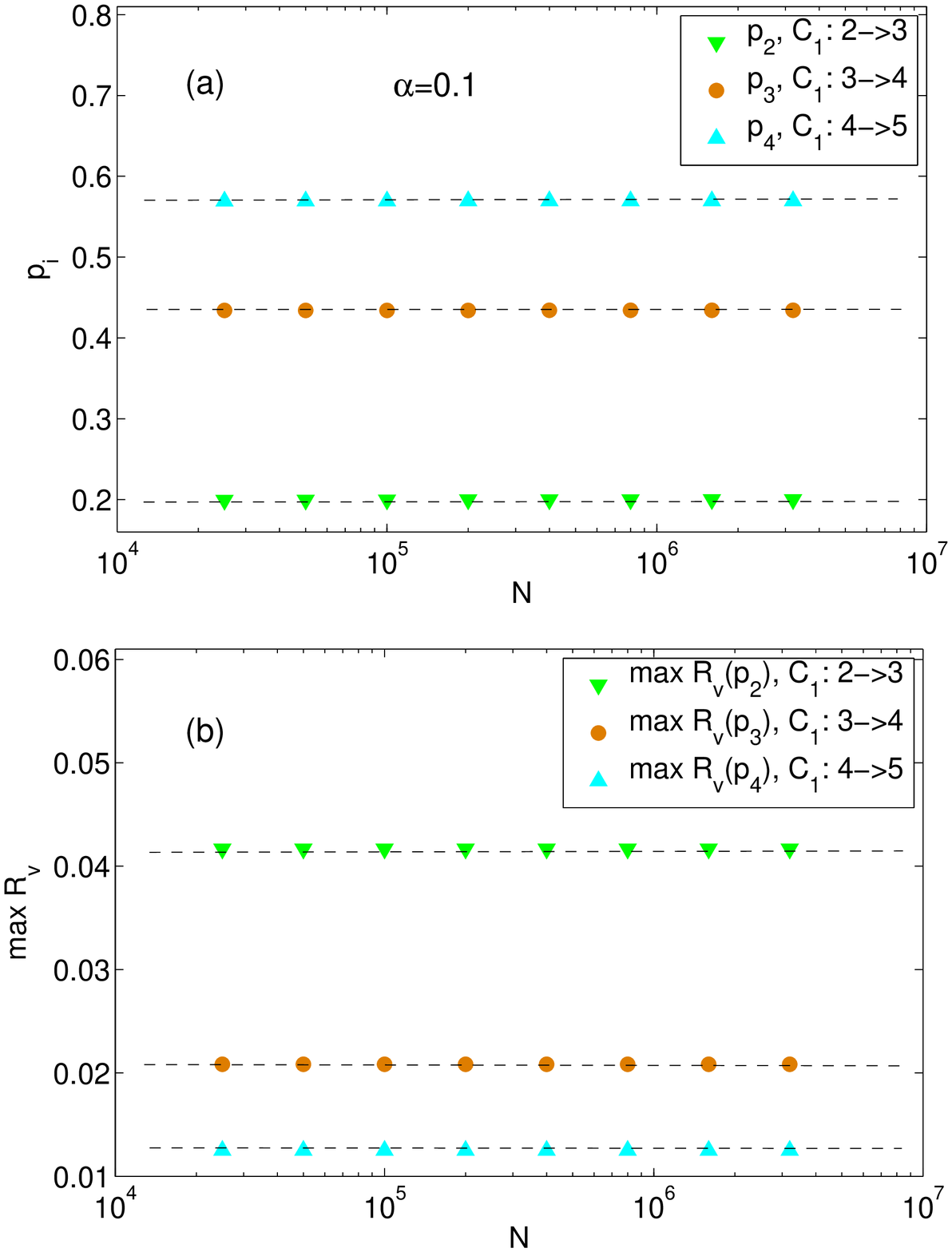}
 \vspace{-0.2in}
\caption{BFW model with $\alpha=0.1$: (a) the location of the $i-th$ peaks of $R_v$, $p_i, i=2,3,4$ is independent of the system size $N$ for BFW model with $\alpha=0.1$. (b) the heights of the $i-th$ peaks of $R_v$, $R_v(p_2), R_v(p_3), R_v(p_4)$ are independent of system size $N$. $R_v(p_i)$ with $i>4$ are also independent of the system size $N$, which are not shown for space constraints.
}
 \label{fig:ds}
 \end{figure}

 In the main manuscript, it was shown that sharp peaks result from micro-micro-transitions $C_1\rightarrow C_1+1$ before the emergence of global connectivity for BFW model with $\alpha=0.6$. Fig.~\ref{fig:as} and Fig.~\ref{fig:bs} show sharp peaks well in advance of $p_c$
for BFW model with $\alpha=0.3$ and $0.1$ respectively, which were also caused by the micro-micro-transitions $C_1\rightarrow C_1+1$. Furthermore, the position of peaks converge to a well-defined set and the peaks do not disappear as the system size increases for BFW model with $\alpha=0.6, 0.3, 0.1$, as Fig.~\ref{fig:BFW2} , Fig.~\ref{fig:cs} and Fig.~\ref{fig:ds} show respectively. 
Fig.~\ref{fig:as} and Fig.~\ref{fig:bs} show that for $\alpha=0.3$ and $0.1$ multiple peaks appear for finite system, however, these multiple peaks shrink into one peak as system size $N\rightarrow\infty$. However, for some values of $\alpha$, multiple peaks survive as $N\rightarrow\infty$ which results from unstable giant components, discussed elsewhere~\cite{chensouzanaglerpaper3}. 

\begin{figure}

 \centering
 \includegraphics[width=0.8\textwidth]{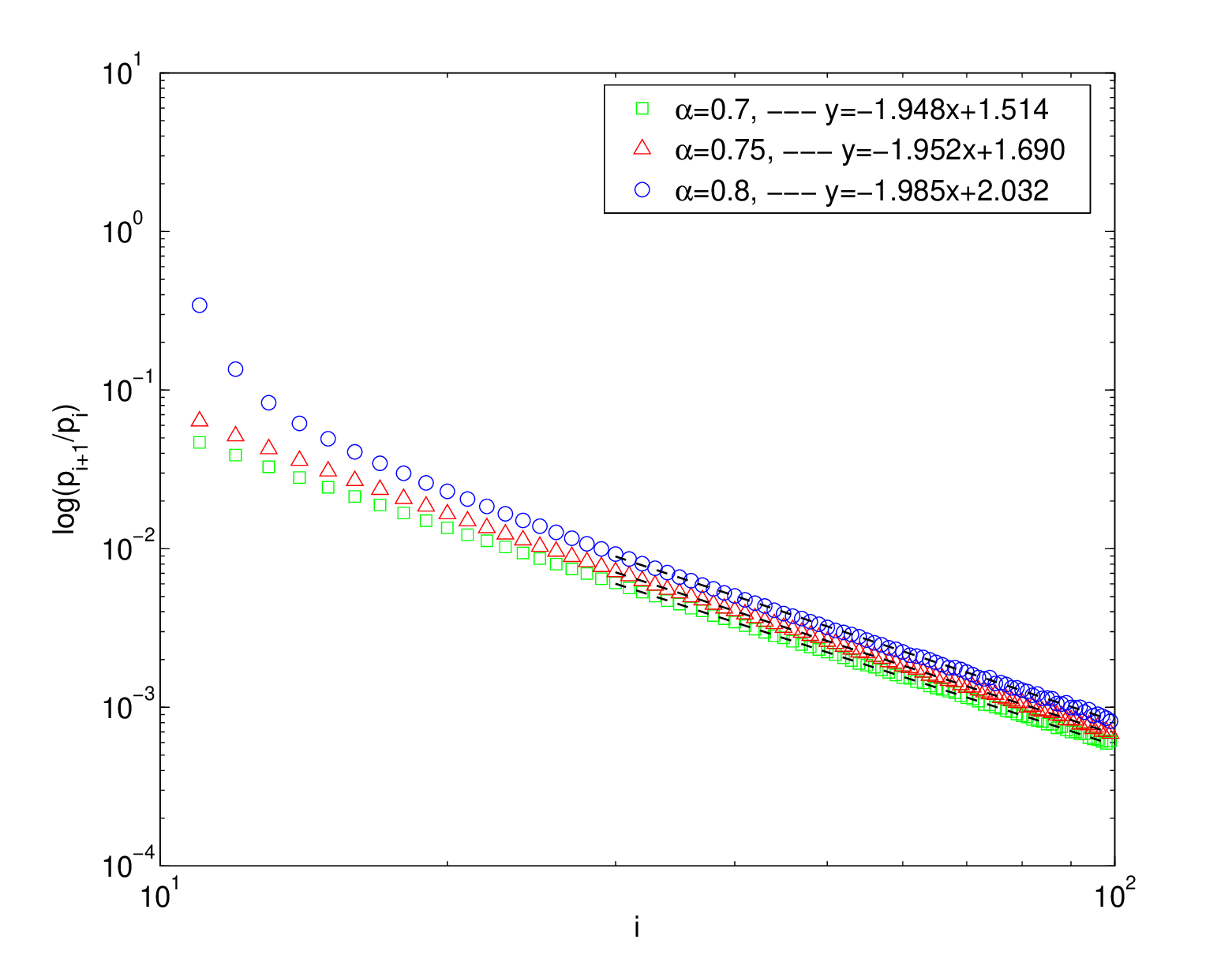}
 \vspace{-0.2in}
\caption{{\bf Scaling Law and Convergence to $p_c$ for the
BFW model} (a) The position $p_i$ of micro-transitions are well fitted by Eqn.~\ref{pis} for $\alpha=0.7, 0.75, 0.8$. Note that we display a log-log plot suggesting $\log\frac{p_{i+1}}{p_i}=Ai^{-b}$.}
 \label{fig:es}
 \end{figure}

From the main manuscript, for the BFW model, the $p_i$ for the jump $C_1: i\rightarrow i+1$ satisfy a recurrence relation of the form
\begin{equation} \label{pis}
 p_{i+1} =\exp( A i^{-b} ) p_i
\end{equation}
with $b$ close to 2, and $A>0$, see fig. 5 (a) in the main manuscript.
Rewriting the recurrence relation Eqn.~(\ref{pis}) as 
\begin{equation} \label{pi0}
p_i = p_{i_0}  \exp\left( A ( (i-1)^{-b}+(i-2)^{-b}+...+i_0^{-b} ) \right) 
\end{equation}
where $i>i_0$.
We set $i_0=50$ in Eqn.~(\ref{pi0}) since the recurrence relation between $p_{i+1}$ and $p_i$ holds for not too small i,
let $i\rightarrow \infty$, we numerically obtain $p_\infty=0.999(1), 0.998(1)$ for $\alpha=0.1, 0.3$ respectively. 
On the other hand, we estimate percolation threshold $p_c$ by measuring the asymptotic location of the largest jump in the order parameter induced by adding a single edge.
We numerically obtain that $p_c=0.999(1), 0.998(1)$ for $\alpha=0.1, 0.3$ respectively, which are equal to $p_\infty$ within error bars. 
We further compare $p_{\infty}$ with $p_c$ for more values of $\alpha>0.6$ where the percolation threshold $p_c$ is significantly smaller than $1$. We numerically obtain that for $\alpha=0.8$, $b\approx1.985, A\approx7.628, p_{i_0}\approx0.72(1), p_{\infty}=0.85(1), p_c=0.86(1)$, for $\alpha=0.75, b\approx1.952, A\approx5.418, p_{i_0}\approx0.76(1), p_{\infty}=0.88(1), p_c=0.89(1)$, and for $\alpha=0.7, b\approx1.948, A\approx4.543, p_{i_0}\approx0.80(1), p_{\infty}=0.90(1), p_c=0.91(1)$ (See Fig.~\ref{fig:es} in detail).
Thus we have $p_\infty$ and $p_c$ are almost equal within error bars for various $\alpha$.

 \section{Derivation of the $p_i$ for the (generalized) BFW model}
 
 
To obtain the values of $p_i$ for a more generalized model, we assume that the form of $g(k)$ in the BFW algorithm is $g(k)=\alpha+(\gamma k)^{-\beta}$ ($\alpha>0, \beta>0, \gamma>0$). Let the probability of sampling an edge which leads to a component of size not larger than $k$ at stage $k$ be $P(k,t,u)$. Begin with stage $k=2$ and at the location of the transition $C_1: 2\rightarrow 3$, about $p_2*N$ edges have been added to the graph since the number of self-edges and multiple edges are of order $o(N)$ and can be omitted. At stage $k=2$, the number of added edges satisfy $0<t\leq p_2*N$ and the probability of sampling an edge which leads to a component of size no larger than $k=2$ is (only an edge that connects two isolated nodes can be added to the system)
\begin{equation}
P(2,t,u)=(1-\frac{2t}{N})^2=(1-2p)^2
\label{p2tu}
\end{equation}
where $p=t/N$.
{ We next analytically obtain $p_2$ for two cases: (i): $g(2)<1$; (ii) $g(2)\geq1$. If $g(2)<1$, from the BFW algorithm, at $p=t/N=p_2$, an edge $e_u$ that leads to a component of size larger than $k=2$ is sampled, since the fraction of accepted edges over randomly sampled edges, $t/u$, satisfies 
\begin{equation}
t/u<g(2),
\label{tug2}
\end{equation}
then $k$ increases from $2$ to $3$ according to BFW algorithm. We can also obtain that 
\begin{equation}
t/(u-1)\geq g(2).
\label{eqngk}
\end{equation}
The reason of above inequality (\ref{eqngk}) is the following. Assume that (\ref{eqngk}) does not hold and $t/(u-1)< g(2)$, then the sampled edge $e_{u-1}$ must lead to a component of size no larger than $k=2$ and it is accepted, resulting in 
\begin{equation}
(t-1)/(u-1)<t/(u-1)<g(2)
\end{equation}
Because otherwise $e_{u-1}$ leads to a component of size larger than $k=2$ and it is rejected, since $t/(u-1)< g(2)$ holds according to our assumption, $k$ would increase from $2$ to $3$ at step $u-1$, which violates our condition that $k$ increase from $2$ to $3$ at step $u$. Then we analyze the situation of the edge sampled at step $e_{u-2}$.
 If this edge leads to a component of size larger than $k=2$ it is rejected.
 Since $t/(u-1)<g(2)$ and $g(2)<1$, we have $(t-1)/(u-2)<g(2)$, so $e_{u-2}$ would lead to $k$ increase from $2$ to $3$ at step $u-2$ which is a contradiction. So we have that $e_{u-2}$ leads to a component of size no larger than $k=2$ and it is accepted, resulting in 
\begin{equation}
(t-2)/(u-2)<(t-1)/(u-1)<g(2).
\end{equation}
With a similar analysis, we recursively obtain that for all steps $m<u$, $e_m$ leads to a component of size no larger than $k=2$ and it is accepted, resulting in 
\begin{equation}
(t-(u-m))/m<g(2)
\label{eqng2}
\end{equation}
Specifically for the step $m=1$ (the first step), from Eq. \ref{eqng2}, we have $t-(u-m)<1$, which is impossible.
 Since the system is initialized with isolated nodes and the first sampled edge would always lead to a component of size no larger than $k=2$ 
 { the edge is accepted,}
 which results in $t-(u-m)=1$ (or $|A|=1$, $A$ is the set of edges accepted or added to the graph) and $m=1$ (number of sampled edges). So we obtain a contradiction which is due to our incorrect assumption that $t/(u-1)<g(2)$. Thus Eq. (\ref{eqngk}) holds.

Combining Eqs. (\ref{tug2}) and (\ref{eqngk}) for $N\rightarrow \infty$, we have following equation at $p=t/N=p_2$,
\begin{equation}
\frac{t/N}{u/N}=g(2).
\label{tnung2t}
\end{equation}

For the case $g(k)\geq 1$, it is much simpler. Since all sampled edges would be accepted by the BFW algorithm, we have
\begin{equation}
\frac{t/N}{u/N}=1.
\label{engo1}
\end{equation}

{
Combining
\begin{equation}
\int \text{d}u/N=\int \frac{\text{d}(t/N)}{P(2,t,u)}
\label{int1}
\end{equation}
with
 $t/N=p_2$ and Eq. (\ref{p2tu}), we have 
$u/N=\int^{p_2}_0 \frac{1}{(1-2p)^2} dp$ as $N\rightarrow \infty$}. 
With 
Eqs. \ref{tnung2t} and \ref{engo1} this results into}
\begin{equation}
\min(1, g(2))*\int^{p_2}_0 \frac{1}{(1-2p)^2} dp=p_2.
\label{min1g2}
\end{equation}
If $g(2)>1$ in Eq. (\ref{min1g2}), since $\frac{1}{(1-2p)^2}>1$, we necessarily have $p_2=0$.
Specifically, we investigate the cases of $\alpha=0.1, 0.3, 0.6, \beta=0.5, \gamma=2$ as studied in our paper.  {Since $g(2)=0.6<1$ for $\alpha=0.1$ and $g(2)=0.8<1$ for $\alpha=0.3$, replace g(2) in Eq. (\ref{min1g2}) and we get $p_2=0.2$ for $\alpha=0.1$ and $p_2=0.1$ for $\alpha=0.3$. However, for $\alpha=0.6$, $g(2)=1.1>1$, we have $p_2=0$. 
{
These results are consistent with observations from our numerical simulations (see figure S4 and S5 in the SI).}

We can also further obtain $p_3$ in a similar way. Firstly, at $p=t/N=p_3$, an edge $e_u$ that leads to a component of size larger than $k=3$ is sampled, and using same analysis as above we have 
\begin{equation}
t/u<g(3)\leq t/(u-1)
\end{equation}
For $N\rightarrow \infty$, we have the following equation at $p=t/N=p_3$
\begin{equation}
\frac{t/N}{u/N}=\min(1, g(3))
\label{ming3}
\end{equation} 
replace $t/N=p_3$ and $u/N=\frac{p_2}{\min(1,g(2))}+\int^{p_3}_{p_2} \frac{1}{P(3,t,u)}dp$ in Eq. (\ref{ming3}), we have 
\begin{equation}
\frac{p_3}{\frac{p_2}{\min(1, g(2))} + \int^{p_3}_{p_2} \frac{1}{P(3,t,u)} dp}=\min(1,g(3))
\label{pkpk1}
\end{equation} 
in which, the form of P(3,t,u) is more complicated than $P(2,t,u)$ and we do not show it here. 
We can further calculate $p_i (i>3)$ recursively with the following relation
\begin{equation}
\frac{p_i}{\frac{p_{i-1}}{\min(1, g(i-1))} + \int^{p_{i}}_{p_{i-1}} \frac{1}{P(i,t,u)} dp}=\min(1,g(i)).
\label{pkpk1}
\end{equation}  
Since $g(i)$ is strictly decreasing, we easily obtain from Eq. (\ref{pkpk1}) that $p_i=p_{i-1}=...=p_2=0$ for $g(i)\geq1$ while $p_i>0$ for $g(i)< 1$}.

\newpage

\begin{figure}
\includegraphics[width=0.8\textwidth]{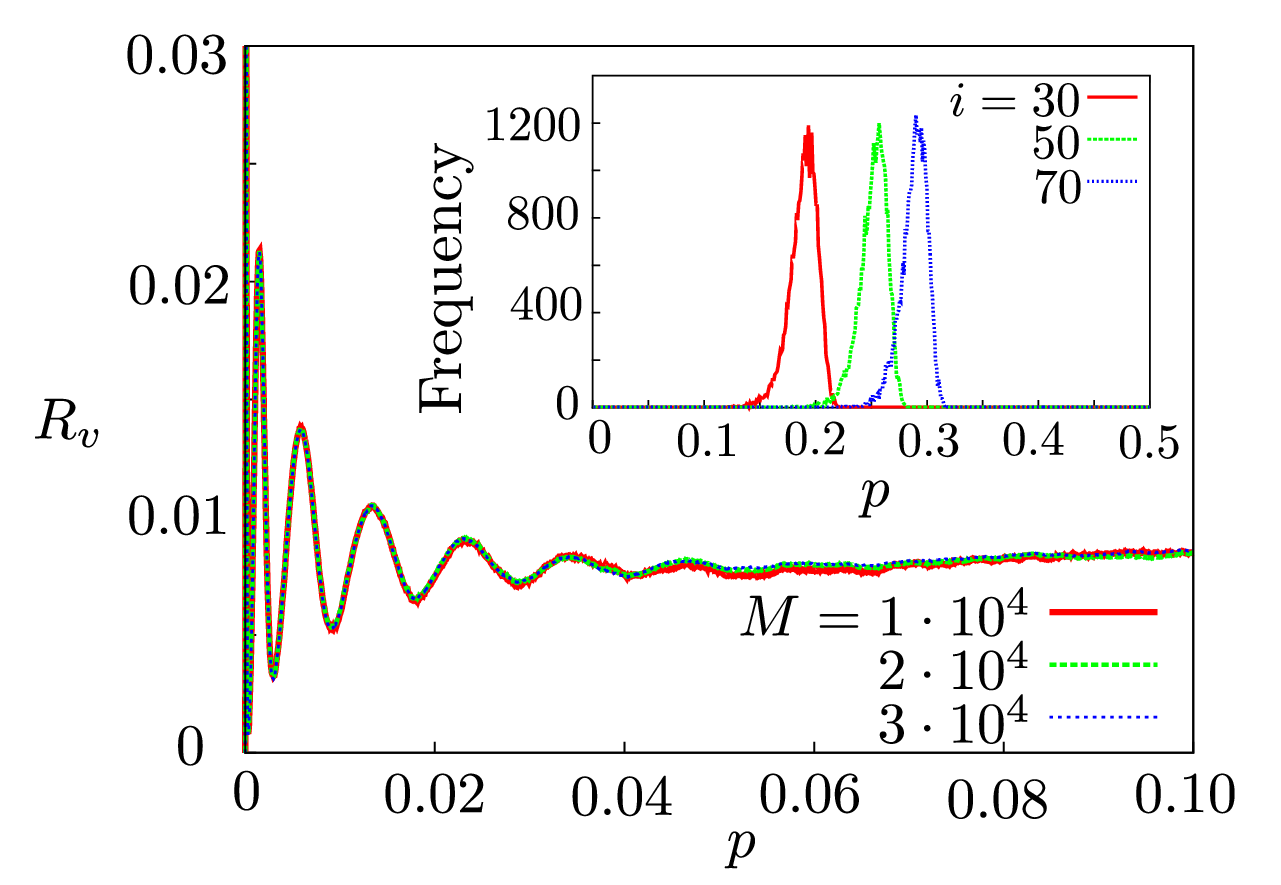}
\caption{ 
\label{fig:newfig}
{\bf Resonances and blurring in $\Rv$ for continuous percolation.}
The relative variance of the order parameter for the ER model of size $N = 2^{25}$ obtained over three
different numbers of realizations $M = 10^4$; $2 \cdot10^4$; $3\cdot 10^4$. The first peak in the plot marks the transition
$C_1: 3 \rightarrow 4$, the second $C_1: 4 \rightarrow 5$ and so on. 
No peaks are seen for jumps for large $i$, due to overlapping broad distributions of the $p_i$, as illustrated in the inset for $i=30, 50, 70$.
}
\end{figure}

   \begin{figure}
 \centering{}
 \includegraphics[width=0.8\textwidth]{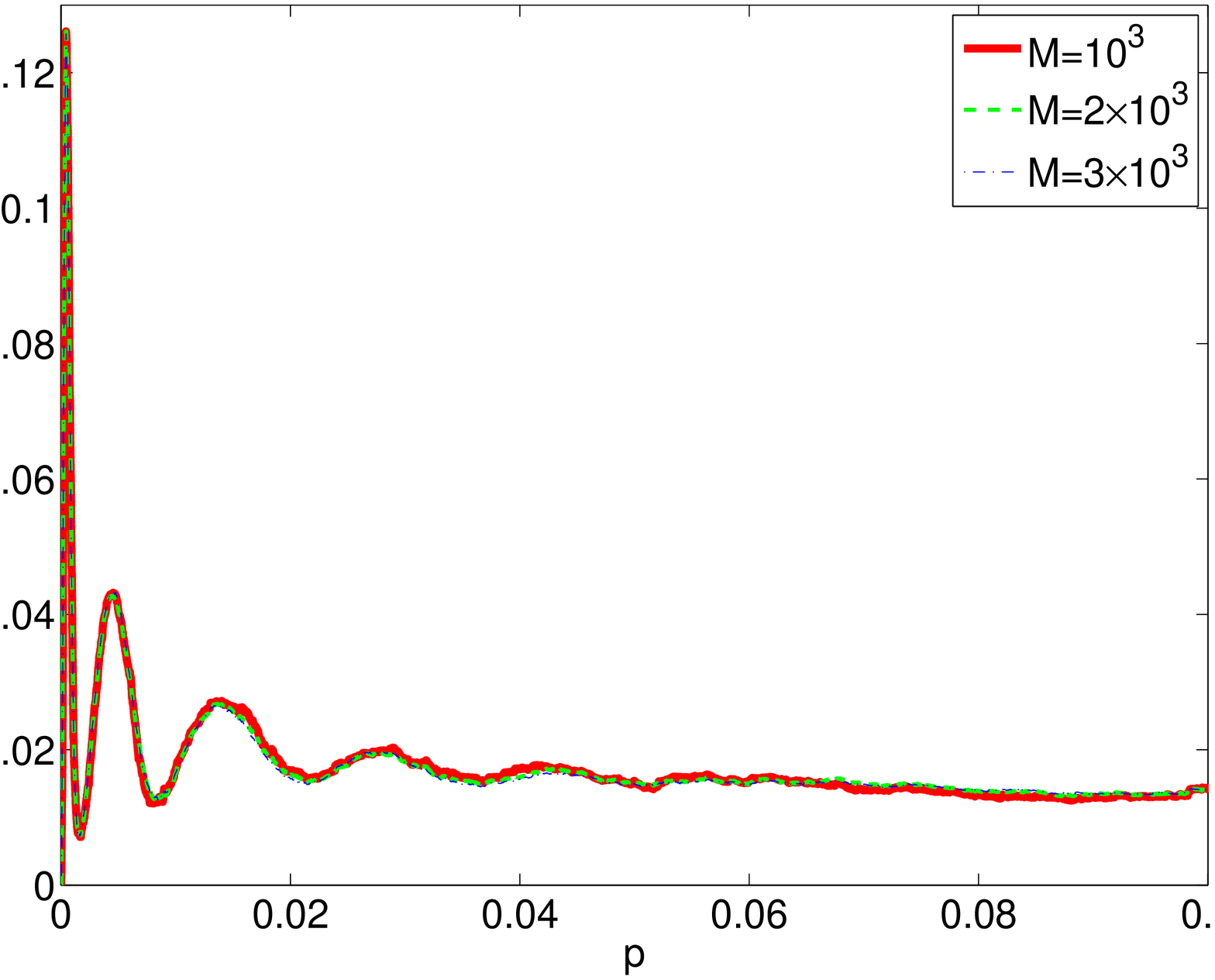}
 \caption{The relative variance of order parameter for 2D lattice percolation of size $N=10^3\times 10^3$ obtained over three different number of realizations $M=10^3, 2\times 10^3, 3\times 10^3$. Remarkably, it shows a qualitatively similar behavior as for ER.}
 \label{fig:malte2}
 \end{figure}
 